\documentclass[11pt]{article}
\usepackage{moriond,epsfig}

\newcommand{\mt}{\ensuremath{m_T}}
\newcommand{\pte}{\ensuremath{p_T^e}}
\newcommand{\vpte}{\ensuremath{\vec{p}_T^{~e}}}

\newcommand{\met}{\ensuremath{{\slash\kern-.7emE}_{T}}}
\newcommand{\metsub}{\ensuremath{{\slash\kern-.5emE}_{T}}}
\newcommand{\vmet}{\ensuremath{\vec{\slash\kern-.7emE}_{T}}}
\newcommand{\ut}{\ensuremath{u_T}}
\newcommand{\vut}{\ensuremath{\vec{u}_T}}
\newcommand{\mee}{\ensuremath{m_{ee}}}
\newcommand{\ptee}{\ensuremath{p_T^{ee}}}
\newcommand{\etaimb}{\ensuremath{\eta_{\rm imb}}}
\newcommand{\upara}{\ensuremath{u_{||}}}

\begin{document}

\mbox{FERMILAB-CONF-12-133-E}

\vspace*{4cm}
\title{Measurement of the $W$ boson mass with 4.3 ${\bf fb^{-1}}$ of D0 Run II data}

\author{Hengne Li\\
(on behalf of the D0 Collaboration)
}

\address{LPSC, Universit\'e Joseph Fourier Grenoble 1, CNRS/IN2P3, \\
   Institut National Polytechnique de Grenoble, Grenoble, France.}

\maketitle\abstracts{
A measurement of the $W$ boson mass using 4.3 fb$^{-1}$ of integrated luminosity collected with the D0 detector during Run II of the Fermilab Tevatron collider is presented. Based on a sample of $1.68\times10^6$ $W\rightarrow e\nu$ candidate events, the $W$ boson mass is measured to be $M_W = 80.367 \pm 0.026$~GeV. Combining this result with an earlier D0 result determined using an independent Run II data sample of 1~fb$^{-1}$ of integrated luminosity, yields $M_W = 80.375 \pm 0.023$~GeV.}

\section{Introduction}

The standard model (SM) of particle physics predicts the existence of a hypothetical scalar particle, the Higgs boson, as a result of the spontaneous electroweak symmetry breaking mechanism that explains the masses of the vector bosons ($W$ and $Z$). Direct searches~\cite{higgssearch} of the SM Higgs boson have limited its possible mass ranges to be 115--127 GeV or above 600 GeV at 95\% C.L.. 

Also predicted by the SM, is a relationship between the $W$ boson mass ($M_W$) and the Higgs boson mass ($M_H$), together with other electroweak parameters such as the top quark mass ($M_t$). Precise knowledge of the value of $M_W$ (and other parameters), therefore, can be used to predict the possible mass range of the hypothetical Higgs boson.

A combination~\cite{b:mwwa} of previous measurements of $M_W$ before winter 2012 yielded a world average value of $M_W = 80.399 \pm 0.023$~GeV. 
This result together with the current $M_t$ measurement~\cite{topmass} (and measurements of other electroweak parameters) predicts~\cite{gruenewald} a mass range of the Higgs boson of $M_H = 92^{+34}_{-26}$~GeV, with an upper limit of 161~GeV at 95\% C.L.. The predicted $M_H$ range overlaps with the ranges allowed by direct searches. However, the predicted range is large, and the experimental precision on the $M_W$ is so far the limiting factor in this prediction. Therefore, improving the precision of the $M_W$ measurements to narrow the predicted $M_H$ range, and comparing the prediction with direct searches to further test the SM, are of great interest. 

This article presents a measurement~\cite{D0IIbW} of $M_W$ using data corresponding to a total integrated luminosity of 4.3~fb$^{-1}$ collected from 2006 to 2009 with the D0 detector~\cite{d0det} at the Fermilab Tevatron $p\bar{p}$ collider.  This measurement uses $W\rightarrow e\nu$ events, with electrons in the central calorimeter (CC, with $|\eta|<1.05$, where $\eta$ is pseudorapidity). The liquid-argon calorimeter of the D0 detector provides stable and accurate electron energy measurements. The energy resolution for an electron in CC at 45~GeV is 4.2\% on average~\footnote{The electron energy resolution depends on the incident angle (or $\eta$) which characterizes the material budget in front of the calorimeter.}.  

The data set used in this analysis has been recorded with increased instantaneous luminosity (almost 3 times higher than in a previous D0 analysis~\cite{D0IIaW}). Higher instantaneous luminosity increases the number of additional $p\bar{p}$ interactions (pile-up) that contaminate the detector. 
They bias the electron energy response
and complicate the modeling of the electron reconstruction efficiency. Therefore, new developments of the analysis techniques are necessary. 


\section{Analysis strategy, event reconstruction}   

We reconstruct two vector variables in the plane transverse to the beam direction from a $W\rightarrow e\nu$ event, namely, the electron transverse momentum (\vpte) and the transverse momentum of the hadronic recoil (\vut) that balances the transverse momentum of the $W$ boson. 

The electron energy is reconstructed as a sum of the energies of calorimeter cells inside the electron reconstruction cone, while the direction of the electron is given by the track in the inner detector that matches spatially to the calorimeter cluster. The electron energy  measurements are corrected for the energy loss due to uninstrumented material in front of the calorimeter. The correction is derived using detailed first-principle simulation. The material budget is determined from a fit to the longitudinal energy profile in the electromagnetic (EM) calorimeter. The gains of the readout cells of the EM calorimeter are calibrated using $Z\rightarrow ee$ events taking the world average $Z$ boson mass~\cite{ZLEP} ($M_Z$) as reference. 

The \vut\ is reconstructed by a vectorial sum of the transverse energies of all the calorimeter cells outside the electron reconstruction cone. The longitudinal component of the hadronic recoil cannot be determined due to the limited pseudorapidity coverage ($|\eta|<4.2$) of the calorimeter. Therefore, the neutrino longitudinal momentum, which is required to reconstruct the invariant mass of the $W$ boson, cannot be determined.

From \vpte\ and \vut, we can calculate three transverse observables for $M_W$ extraction: the transverse mass of the $W$ boson ($\mt = \sqrt{(\pte+\met)^2-(\vpte+\vmet)^2}$), the electron transverse momentum ($\pte = |\vpte|$), and the missing transverse energy ($\met = |- \vpte - \vut |$) due to the neutrino transverse momentum. 

A fast Monte-Carlo (MC) model is developed to generate a series of templates for the above three observables based on different $M_W$ hypotheses.  The $M_W$ is determined, specially for each observables, using a binned likelihood fit of the predicted templates to the data.

For the $W\rightarrow e\nu$ event selection, we require an electron in CC with $\pte>25$~GeV. The event is required to satisfy $\met>25$~GeV, $\ut<15$~GeV, and $50<\mt<200$~GeV. The requirement on $\ut$ is made to constrain the transverse boost of the $W$ boson, since the transverse boost of the $W$ boson degrades the sharpness of the Jacobian edge in the $\pte$ distribution. However, this treatment also translates certain uncertainties from the hadronic recoil modeling to the $\pte$. There are 1\,677\,394 candidate $W\to e\nu$ events after selection.

The $Z\rightarrow ee$ events are the control sample for tuning the fast MC, such as the electron energy scale and the hadronic recoil model. The $Z\rightarrow ee$ events are selected by requiring two electrons both with $\pte>25$~GeV. Events are also required to have $\ut<15$~GeV to constrain the transverse boost of the $Z$ boson, and $70<\mee<110$~GeV, where $\mee$ is the invariant mass of the electron pair. There are 54\,512 $Z\rightarrow ee$ candidate events with both electrons in CC, which are used for most of the model tuning. Events allowing one electron in the end calorimeter (EC, with $1.5<|\eta|<2.5$) are only used for measurements of the electron reconstruction efficiency.

\section{Fast Monte-Carlo model}

The fast Monte-Carlo (MC) model for template generation has to simulate $W$ and $Z$ boson production and decay, the electron energy response, the hadronic recoil, the underlying events contamination, the electron reconstruction efficiency, and the background.

\subsection{Boson production and decay}

The boson production and decay are simulated using {\tt RESBOS}~\cite{resbos} event generator combined with {\tt PHOTOS}~\cite{photos}. {\tt RESBOS} is a next-to-leading order event generator including next-to-next-to-leading order logarithm resummation of soft gluons. {\tt PHOTOS} generates up to two final state radiation (FSR) photons. Parton distribution functions are described using CTEQ6.6~\cite{cteq66}. The boson transverse momentum prediction in {\tt RESBOS} is dominantly determined by the nonperturbative parameter~\cite{g2} $g_2$. The $g_2$ value~\cite{d0g2} $0.68\pm0.02~{\rm GeV}^2$ is used.

\subsection{Electron energy response}

The electron energy response is modeled by firstly modeling the energy responses that are not a linear function of the electron true energy. Then, we assume the rest of the energy response is a linear function of the electron true energy, fit to the $Z\rightarrow ee$ data sample to determine the scale. 

The energy loss correction, as one of the non-linear energy responses, is applied to the data. There are also certain non-linear energy responses due to the high instantaneous luminosity to be modeled in the fast MC.  

One of them is the reduction of EM calorimeter gain due to a high voltage (HV) drop caused by a large instantaneous pile-up energy deposition.  A large current that flows through the resistive coat of the HV pads of calorimeter cells creates a reduction of the HV. The HV supplies are connected at both ends of the CC modules (at $|\eta|=1.2$). Therefore, HV drop is larger for cells at small $|\eta|$ than for cells at large $|\eta|$. This gain loss is modeled as function of instantaneous luminosity and detector $\eta$ in the fast MC. The EM calorimeter calibration at the cell level applied to the data is done in the absence of the knowledge of the HV drop. Certain imperfections in such a calibration are expected. Thus, an addition model of the residual miscalibration as a function of detector $\eta$ is introduced in the fast MC. 

Another non-linear energy response is understood as an effect of the electron reconstruction cone. The electron energy is reconstructed as a sum of energies deposited in a cone consisting of 13 calorimeter towers.  Not only the electron deposits its energy into this cone, but also some of the hadronic recoil, pile-up, and spectator parton interactions. The energy deposition from the latter sources does not come from the true electron but is reconstructed as part of the electron energy in the data. This additional energy contribution is modeled in the fast MC as a function of instantaneous luminosity, $\eta$, $\upara$ (the $\vut$ projection to the electron direction), and $SET$ (the scalar sum of the transverse energy deposited all over the calorimeter with cells in the electron reconstruction cone excluded).

\begin{figure}[hbpt]
  \centering
  \includegraphics[width=0.32\linewidth]{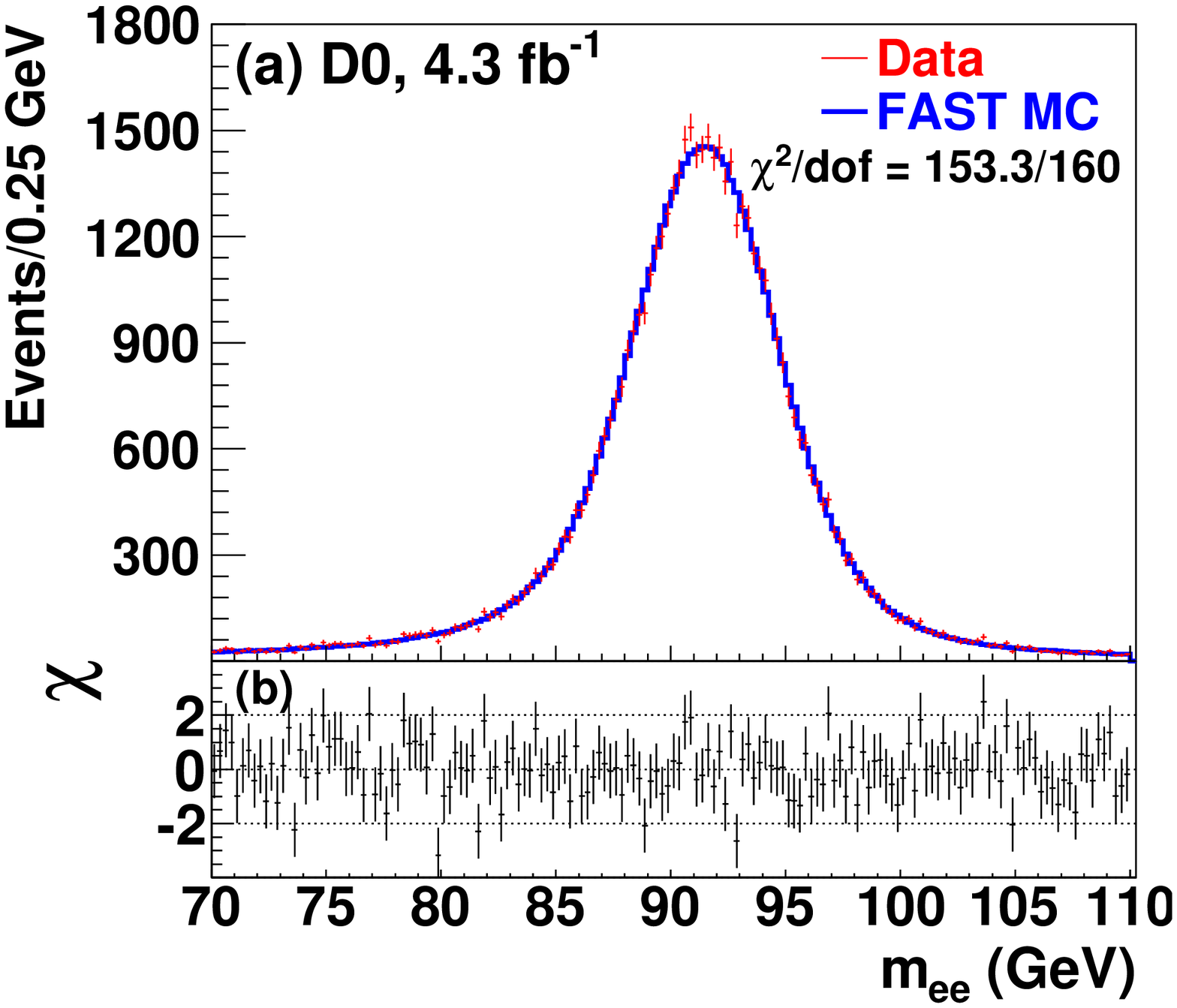}
  \includegraphics[width=0.42\linewidth]{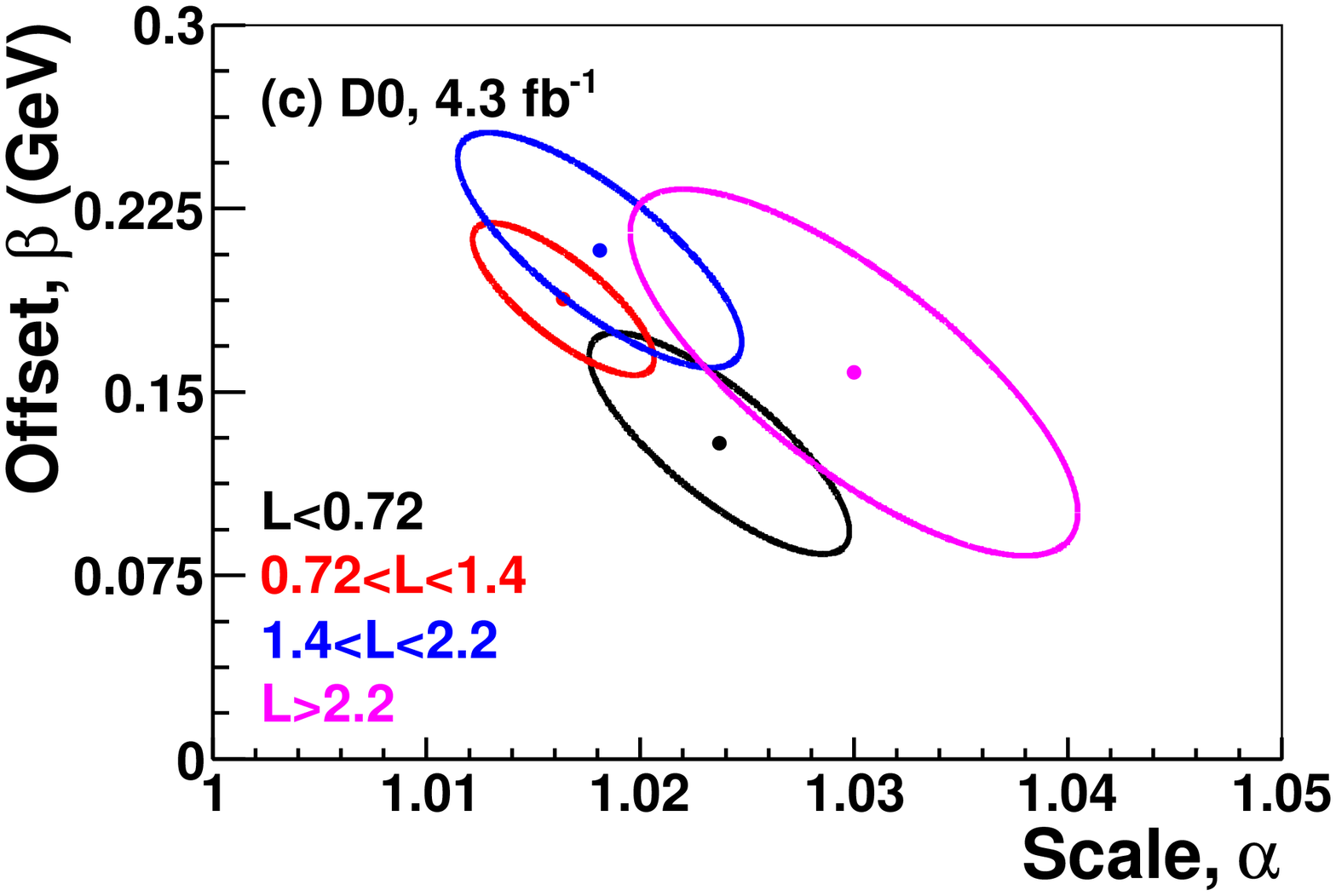}
  \caption{(a) The dielectron invariant mass distribution in $Z\to ee$ data and 
    from the fast MC, (b) the $\chi$ plot of (a), and (c) the fitted scale and offset 1-sigma contours in bins of instantaneous luminosity (in units of $10^{32}{\rm cm}^{-1}{\rm s}^{-1}$).  
    \label{f:zfinal}}
\end{figure}

After modeling of the non-linear responses, the linear response is modeled as 
$E = \alpha \cdot (E_{true}-43~{\rm GeV}) + \beta + 43~{\rm GeV}$,
where, $\alpha$ is the energy scale, $\beta$ is the energy offset, and 43~GeV is an arbitrary offset introduced technically to improve the stability of the fit for $\alpha$ and $\beta$. The parameters $\alpha$ and $\beta$ are determined by a template fit to the $\mee$ versus $f_Z$ distribution of the $Z\rightarrow ee$ events, where $f_Z = (E_1+E_2)\cdot (1-\cos{\omega}) / \mee$,  $E_1$ and $E_2$ are energies of the two electrons, and $\omega$ is the opening angle between the two electrons. The $\alpha$ and $\beta$ are determined separately for four instantaneous luminosity sub-samples, and are consistent with each other, as shown in Figure~\ref{f:zfinal}~(c). After the electron energy scale tuning, a $Z$ boson mass fit returns $M_Z=91.193\pm0.017{\rm (stat)}$~GeV, which is in good agreement with the world average ($M_Z=91.188$~GeV). The $M_Z$ fit is shown in Figure~\ref{f:zfinal}~(a).

\subsection{Hadronic recoil}

The hadronic recoil ($\vut$) reconstructed from the data contains contributions from the hadrons recoiling against the $W$ boson, and additional components that are (mostly) independent of the $W$ boson boost. These additional components include spectator parton interactions, pile-up, a small part of all the components above that enter the electron reconstruction cone, and the FSR photons outside the electron reconstruction cone. 

All these components are modeled separately in the fast MC, and a vectorial sum of them gives the $\vut$. Free parameters are reserved to tune the recoil model. The hadronic response and resolution are tuned using the mean and width of the $\etaimb$ distributions, respectively, of the $Z\rightarrow ee$ events in bins of $\ptee$, where $\etaimb$ denotes the projection of the sum of the dielectron transverse momentum and $\vut$ vectors on the axis bisecting the dielectron directions in the transverse plane~\cite{ua2eta}.

\subsection{Efficiency}

The kinematic dependence of the electron reconstruction efficiency sculpts the distributions of the observables used to measure the $M_W$. The pile-up due to high instantaneous luminosity and the hadronic recoil are the two major sources that contaminate the electron reconstruction window and give inefficiency in the electron reconstruction. The effects of pile-up can indirectly introduce kinematic dependence of the efficiency. For instance, under a given pile-up contamination, a high energy electron can be more easily identified than a low energy electron. The hadronic recoil contamination depends on the amount of hadronic activity and also the relative orientation of the hadronic recoil with respect to the electron. 

The kinematic dependence of the efficiency is determined and modeled in the fast MC in two steps. In the first step, we extract the efficiency dependence from a high statistics detailed {\tt GEANT}~\cite{b:geant} MC simulation (full MC) of the $W\rightarrow e\nu$ and $Z\rightarrow ee$ events generated using the {\tt PYTHIA}~\cite{pythia} event generator, and model it in the fast MC. The full MC is overlaid at the cell level with a dedicated pile-up collider data sample which is weighted according to the instantaneous luminosity distribution (with random bunch crossings) of the data set used in this analysis.  This step simplifies the modeling of the complicated correlations among the various efficiency dependencies with a data-based determination of the pile-up impacts. The second step is to extract the dependence of the efficiency on the major variables ($\pte$, $\upara$, $SET$, instantaneous luminosity, etc.) from the data, and compare them with those from the full MC. Excellent agreement is found between the full MC and collider data.

\subsection{Backgrounds}

Backgrounds in the $W$ boson candidate sample modify the shapes of the distributions of the three observables. The major backgrounds are $Z\rightarrow ee$ events where one of the electrons escapes detection, multijet events where a jet is misidentified as an electron with $\met$ arising from misreconstruction, and $W \rightarrow \tau \nu \rightarrow e\nu\nu\nu$ events. The fractions of the backgrounds in the $W$ boson candidate sample are 1.08\% for $Z\rightarrow ee$, 1.02\% for multijet events, and 1.67\% for $W \rightarrow \tau \nu \rightarrow e\nu\nu\nu$ events. The impact of the uncertainties in the background model on the $M_W$ measurement is found to be small.

\section{Results and outlook}

\subsection{Systematic uncertainties}

The systematic uncertainties of the $M_W$ measurements are listed in Table~\ref{t:syst}. They are divided into two categories: from experimental sources and from boson production and decay modeling.

Among the experimental aspects, the uncertainties from electron energy calibration, electron energy resolution model, and hadronic recoil model are driven by the limited statistics of the $Z\rightarrow ee$ control sample. The shower modeling systematic uncertainties reflect the uncertainties in the amount of uninstrumented material, and the energy loss systematic uncertainties arise from the finite precision of the simulation of electron showers based on a detailed model of the detector geometry. The systematic uncertainties of electron calibration, electron resolution,  electron reconstruction efficiency, hadronic recoil model and backgrounds are determined by varying the corresponding parameters within the statistical uncertainties of their measurements. 

The uncertainties due to boson production and decay modeling are dominantly due to the PDFs. In principle, the transverse observables used in the $M_W$ measurement are insensitive to the uncertainties of the (longitudinal) PDFs. However, our requirements on the lepton pseudorapidity ($|\eta|<1.05$) is not invariant  under longitudinal boosts. Changes in the PDFs can modify the shapes of the transverse observables in the presence of the pseudorapidity requirements. The PDF uncertainties are propagated to $M_W$ by generating ensembles of $W$ boson events using {\tt PYTHIA} with CTEQ6.1~\cite{cteq61}. The QED uncertainties are estimated by comparing {\tt PHOTOS} to {\tt WGRAD}~\cite{wgrad} and {\tt ZGRAD}~\cite{wgrad} event generators, which provide a more complete treatment of electroweak corrections at the one radiated photon level. The uncertainties from boson transverse momentum modeling is determined by propagation of the uncertainty of the $g_2$ parameter.

\begin{table}[ht]
\begin{center}
  \caption{Systematic uncertainties of the $M_W$  measurement.\label{t:syst}}
  \begin{tabular}{|l|ccc|}
  \hline
          & \multicolumn{3}{c|}{$\Delta M_W$~(MeV)} \\
   Source                          &$\mt$ & $\pte$ &  $\met$\\
  \hline 
  Electron energy calibration       & 16 &  17 & 16 \\
  Electron resolution model         &  2 &   2 &  3 \\
  Electron shower modeling           &  4 &   6 &  7 \\
  Electron energy loss model        &  4 &   4 &  4 \\
  Hadronic recoil model             &  5 &  6 & 14 \\
  Electron efficiencies             &  1 &   3 &  5 \\
  Backgrounds                       &  2 &   2 &  2 \\ \hline
  Experimental subtotal             & 18 &  20 & 24 \\ \hline
				    				     
  PDF                          &  11 &  11 & 14 \\
  QED                          &  7 &   7 &  9 \\
  Boson $p_T$                  &  2 &   5 &  2 \\ \hline
  Production subtotal          & 13 &  14 & 17 \\ \hline

  Total                        &  22 & 24 & 29 \\ \hline
  \end{tabular}
\end{center}
\end{table}

\subsection{Results}

\begin{figure}[hb]
  \includegraphics[width=0.32\linewidth]{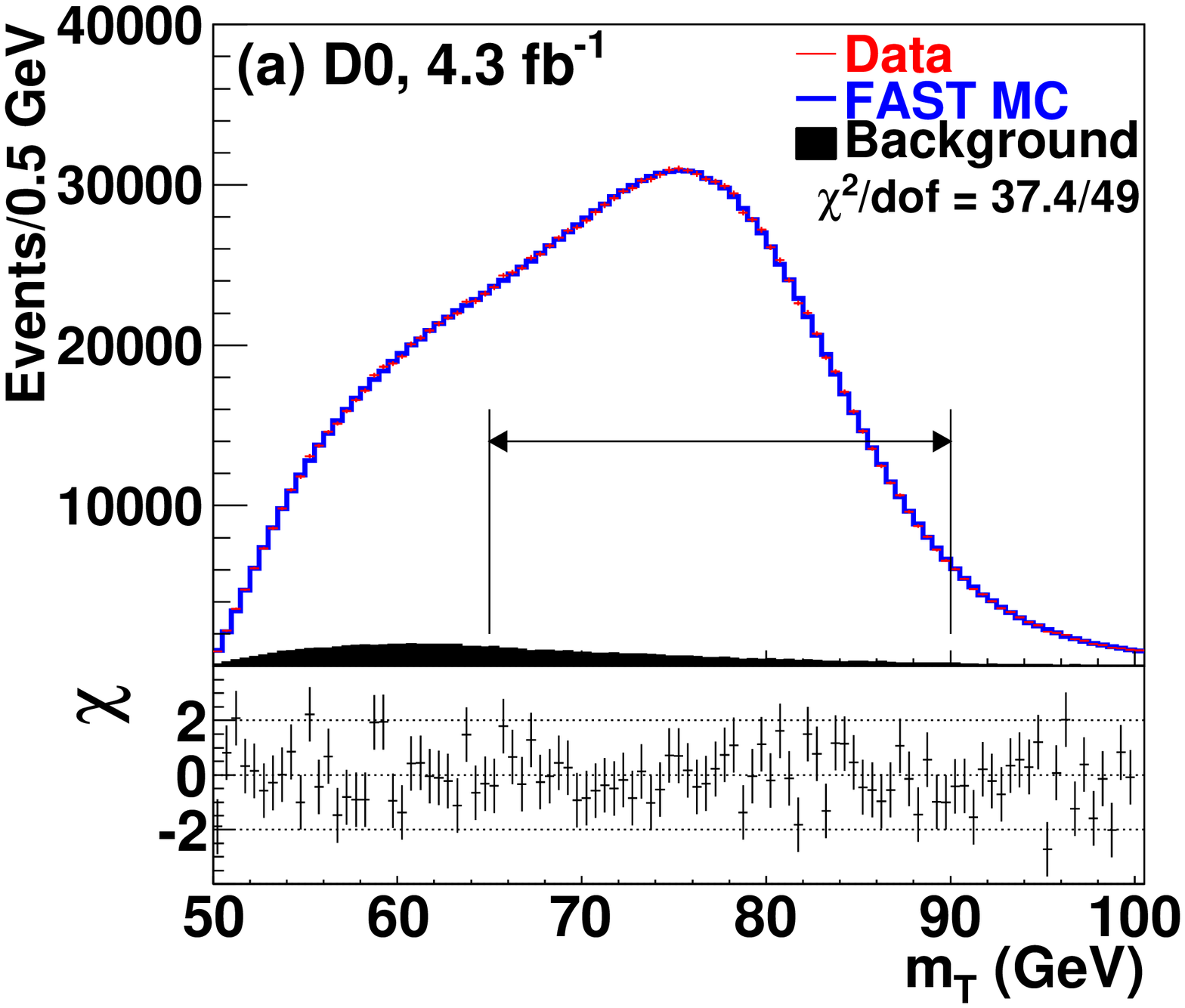}
  \includegraphics[width=0.32\linewidth]{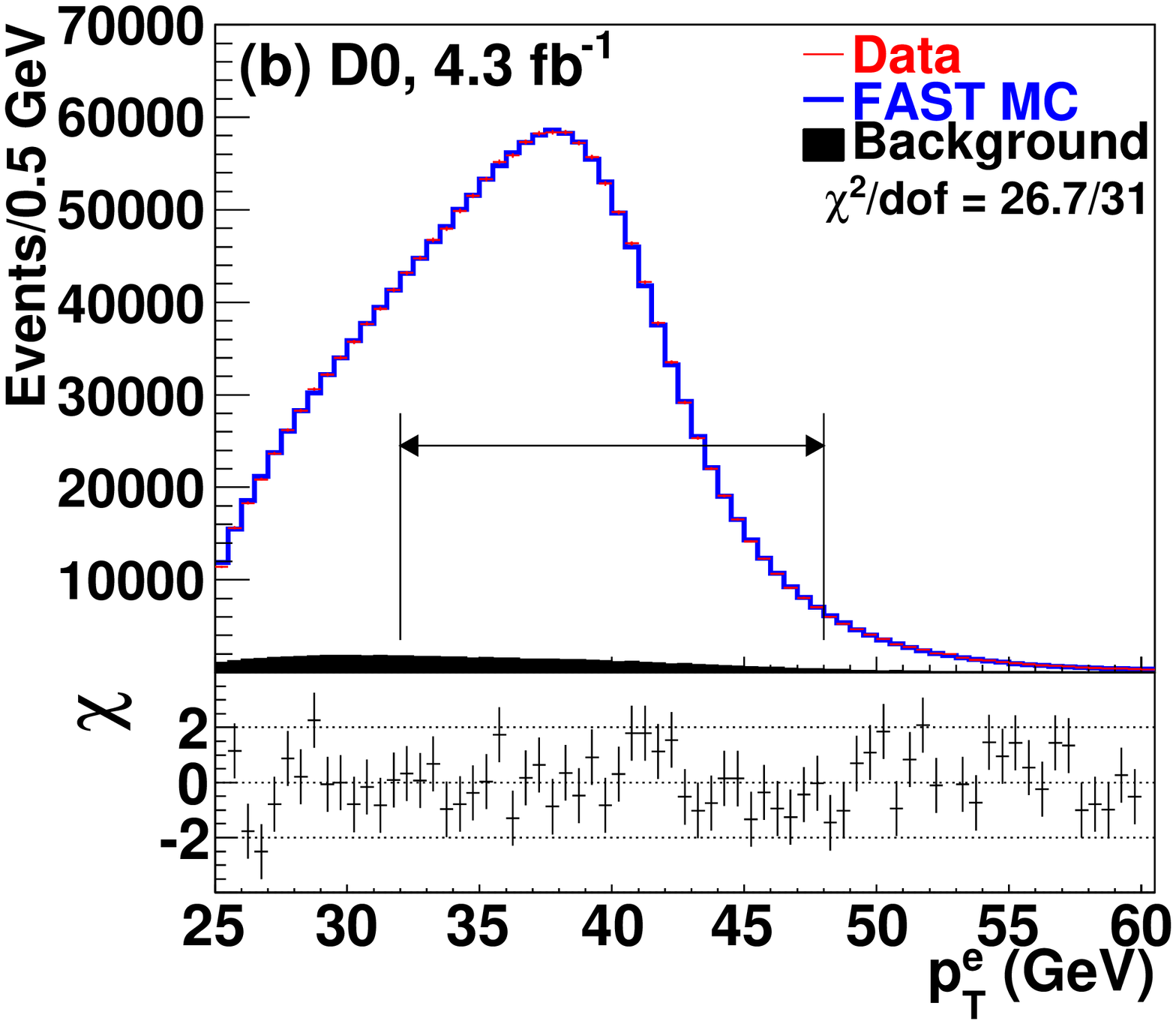}
  \includegraphics[width=0.32\linewidth]{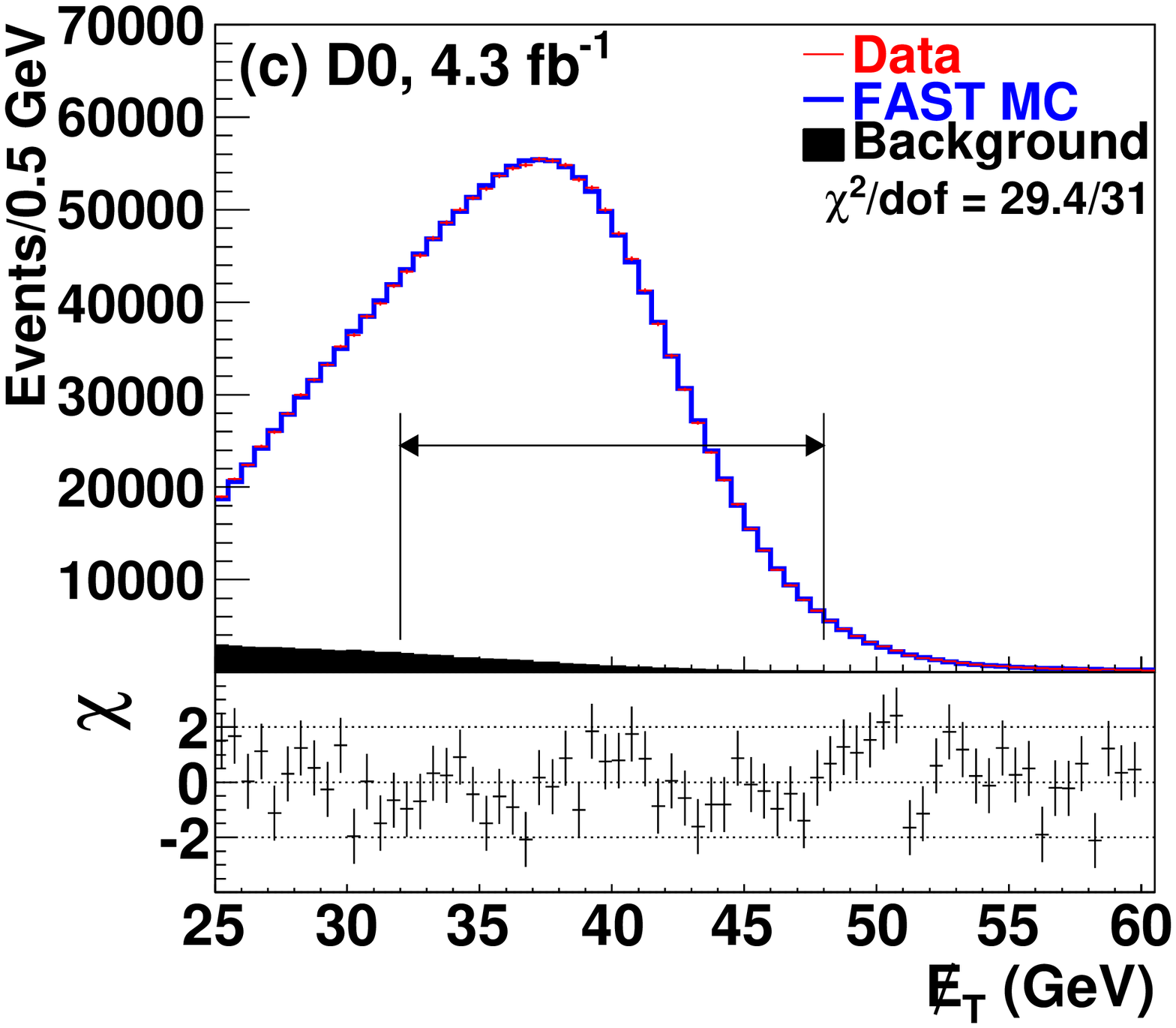}
  \caption{The (a) $\mt$, (b) $\pte$, and (c) $\met$ distributions for data and
   fast MC simulation with backgrounds, together with the $\chi$ plots comparing data and fast MC.
    The fit  ranges are indicated by the double-ended horizontal arrows.\label{fig:fits}}
\end{figure}

The value of $M_W$ is extracted by fitting templates of the three observables ($\mt$, $\pte$, and $\met$) generated by the fast MC to the distributions from the collider data. The fitting results are shown in Table~\ref{tab:fits}, together with optimized fit ranges, for the three observables. Figure~\ref{fig:fits} shows the distributions of the three observables in data and the comparison with templates from fast MC for the best fit $M_W$.  During the tuning of the fast MC to describe the collider data, an unknown constant offset is added to the $M_W$ values returned from the fits, which is the same for the three observables. This enables the tuning of the fast MC to be done without the knowledge of the final results. 
\begin{table}[hbtp]
\begin{center}
  \caption{Results from the fits to data. The quoted uncertainty is solely due to the statistics of the $W$ boson sample.  \label{tab:fits}}
  \begin{tabular}{|cccc|}
     \hline
     Variable & Fit Range (GeV) & $M_W$ (GeV)     & $\ \ \ \chi^2$/dof \\ \hline
      $\mt$   & $65<\mt<90$      & $\ \ \ 80.371\pm0.013\ \ \ $ &  37.4/49  \\
     $\pte$   & $32<\pte<48$      & $      80.343\pm0.014      $ &  26.7/31  \\ 
     $\met$   & $32<\met<48$    & $      80.355\pm0.015      $ &  29.4/31  \\
     \hline
  \end{tabular}
\end{center}
\end{table}

Combining the results from $\mt$ and $\pte$ methods using the BLUE~\cite{blue} method,  we obtain the final result of the 4.3~fb$^{-1}$ measurement:
\begin{eqnarray*}
  M_W & = & 80.367 \pm 0.013\ \mathrm{(stat.)} \pm 0.022\ \mathrm{(syst.)\ GeV}\\
      & = & 80.367 \pm 0.026\ \mathrm{GeV}.
\end{eqnarray*}
The result from the $\met$ method is not used in the combination. In the combination, we assume 100\% correlation for those uncertainties that are nonstatistical in nature, such as the QED uncertainties, to protect them from being decreased. However, with this protection, the BLUE combination gives a sizable negative weight for the $M_W$ value from the $\met$ method which has relatively larger uncertainties. The interpretation is that the central values of $M_W$ from the other more precise methods ($\mt$ and $\pte$) have also fluctuated apart from the true value of $M_W$ in the same direction as the less precise $\met$ method. Given that it is our protection that introduces the negative weight, and the contribution to the combined precision from the $\met$ method is negligible, we decide to only use the $\mt$ and $\pte$ methods in the above combination to avoid the potential bias.

We combine our result with the earlier D0 measurement~\cite{D0IIaW} to obtain the D0 5.3~fb$^{-1}$ result:  
\begin{eqnarray*}
  M_W & = & 80.375 \pm 0.011\ \mathrm{(stat.)} \pm 0.020\ \mathrm{(syst.)\ GeV}\\
      & = & 80.375 \pm 0.023\ \mathrm{GeV}.
\end{eqnarray*}
The precision achieved is the same as the previous world average.

\begin{figure}[hbtp]
\centering \includegraphics [width=0.36\textwidth]{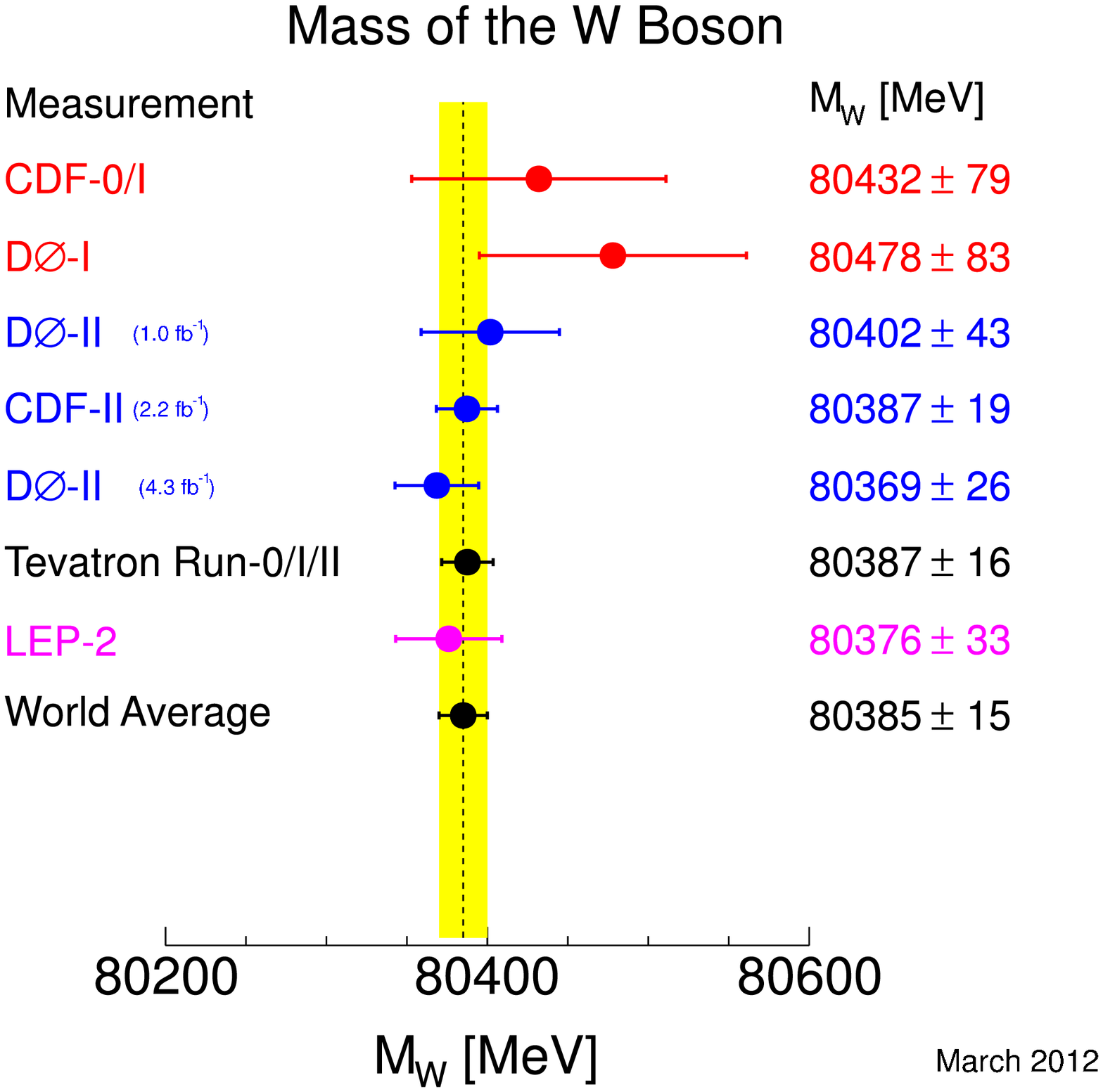}
\centering \includegraphics [width=0.36\textwidth]{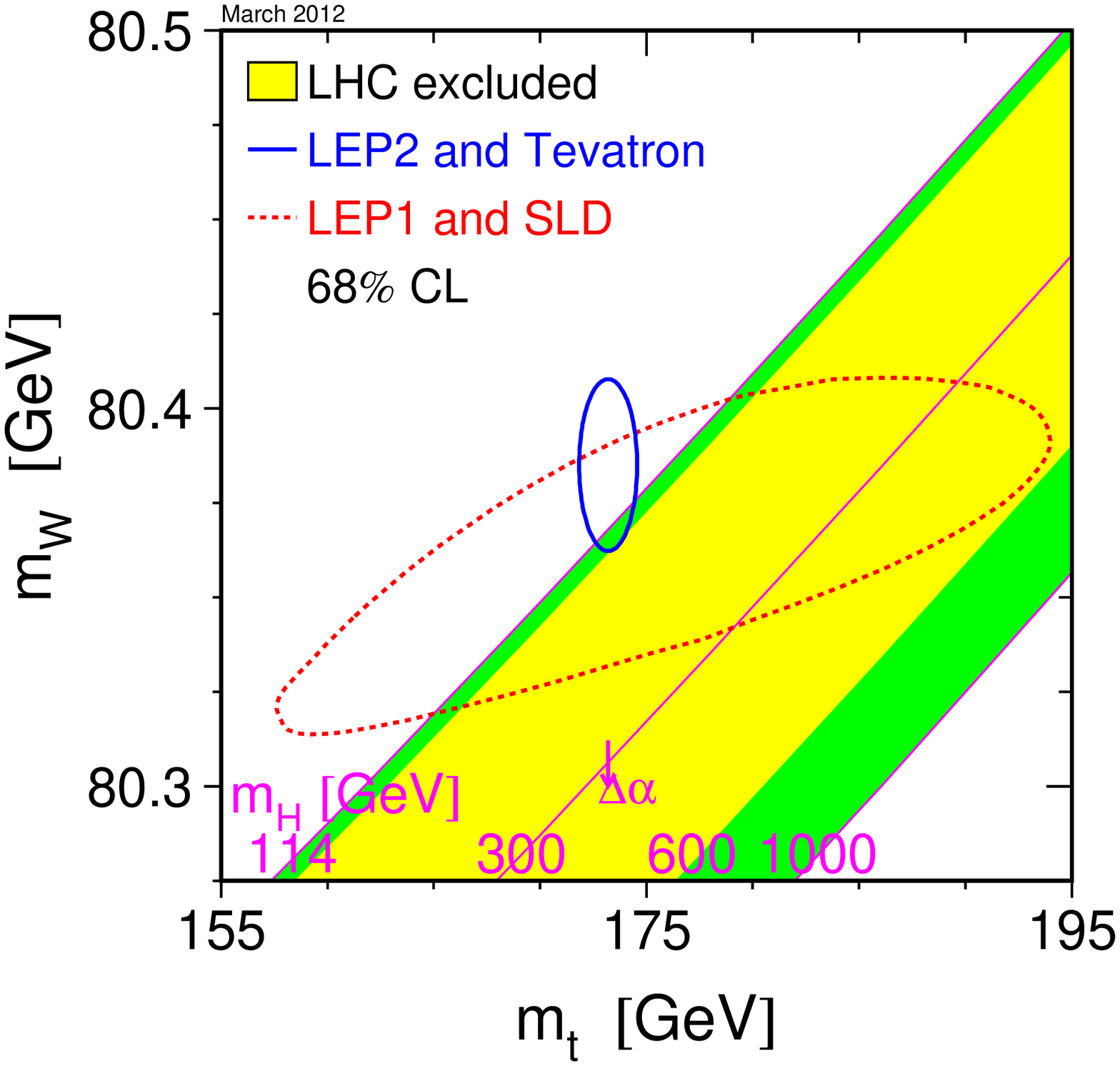}
\caption{(left) Summary of the measurements of the $W$ boson mass and their average as of March 2012. (right) 1-sigma contour in the ($M_t$, $M_W$) plane representing the current world averages of $M_t$ and $M_W$ measurements. The green bands show the possible Higgs boson masses that are not excluded by direct searches.}
\label{fig:sum}
\end{figure}

The combination with all previous measurements and the recent CDF measurement~\cite{CDFIIbW} gives the new world average~\cite{Comb2012}: 
\begin{eqnarray*}
  M_W & = & 80.385 \pm 0.015\ {\rm GeV}.
\end{eqnarray*}
The indirect constraints~\cite{gruenewald} on the Higgs boson mass based on this new world average is  $M_H = 94^{+29}_{-24}$~GeV, with an upper limit of 152~GeV at 95\% C.L.. A summary~\cite{Comb2012} of the measurements of the $M_W$ and their average is shown on Figure~\ref{fig:sum}~(left). The world averages of $M_W$ and $M_t$ measurements with still allowed regions after direct searches of the Higgs boson are shown in Figure~\ref{fig:sum}~(right)~\cite{gruenewald}.  

\subsection{Outlook}

D0 has another $\sim5$~fb$^{-1}$ collider data to be analyzed. Including this last data set and still only using electrons in CC, if we assume the uncertainties from boson production and decay model would be unchanged, the precision of $M_W$ from the total $\sim10$~fb$^{-1}$ D0 data is expected to be $\sim18$~MeV. 

If the PDF uncertainties would be reduced by a factor of two by using new PDF sets (e.g. Ref.~\cite{ct10w}) including constraints from $W$ charge asymmetry measurements,  and also using electrons in EC to extend the pseudorapidity coverage, the above precision can be improved to $\sim16$~MeV.


\end{document}